\documentclass[prb, twocolumn, showpacs,preprintnumbers,superscriptaddress]{revtex4}

\usepackage{amsfonts}
\usepackage{amsmath}
\usepackage{amssymb}
\usepackage{graphicx}

\begin{document}

\title{Theoretical modeling of spatial and temperature dependent exciton energy in coupled
quantum wells}

\author{C. S. Liu}
\affiliation{Department of Physics, Yanshan University, Qinhuangdao
066004, China} \affiliation{Department of Physics, National Taiwan
Normal University, Taipei 11650, Taiwan}

\author{H. G. Luo}
\affiliation{Center for Interdisciplinary Studies, Lanzhou
University, Lanzhou 730000, China} \affiliation{Key Laboratory for
Magnetism and Magnetic Materials of the Ministry of Education,
Lanzhou University, Lanzhou 730000, China} \affiliation{Institute of
Theoretical Physics, Chinese Academy of Sciences, Beijing 100080,
China}

\author{W. C. Wu}
\affiliation{Department of Physics, National Taiwan Normal
University, Taipei 11650, Taiwan}

\date{\today}

\begin{abstract} Motivated by a recent experiment
of spatial and temperature dependent average exciton energy
distribution in coupled quantum wells [S. Yang \textit{et al.},
Phys. Rev. B \textbf{75}, 033311 (2007)], we investigate the nature
of the interactions in indirect excitons. Based on the uncertainty
principle, along with a temperature and energy dependent
distribution which includes both population and recombination
effects, we show that the interplay between an attractive two-body
interaction and a repulsive three-body interaction can lead to a
natural and good account for the nonmonotonic temperature dependence
of the average exciton energy. Moreover, exciton energy maxima are
shown to locate at the brightest regions, in agreement with the recent
experiments. Our results provide an alternative way for
understanding the underlying physics of the exciton dynamics in
coupled quantum wells.
\end{abstract}



\maketitle

\section{Introduction}

An exciton is a bound pair of an electron and a hole in a
semiconductor. At low densities, exciton are Bose quasiparticles
with a small mass, somewhat similar to hydrogen atoms. At low
temperatures, of the order of 1 Kelvin, the excitons are expected to
realize the phenomenon of Bose-Einstein condensation (BEC).
\cite{Keldysh1968} Recently, it has been shown that indirect exciton
(spatially separated electron-hole pairs) in coupled quantum wells
(CQW) can have a long lifetime and high cooling rate. With these
merits, Butov \textit{et al.} have successfully cooled the trapped
excitons to the order of 1 K. \cite{Butov2002a} Although there is not
enough evidence to prove that these excitons are condensed into the
BEC state, it is fascinating enough to observe several novel
features of the photoluminescence (PL) patterns.\cite{Butov2002b,
LaiCW2004}

Major features of the macroscopically ordered exciton states
observed are summarized as follows.\cite{Butov2002b, LaiCW2004} (i)
Two exciton rings are formed. When the focused laser is used to
excite the sample and prompt luminescence is measured in the
vicinity of the laser spot, a ring, called the internal ring, is
formed. A second ring of PL appears away from the source at the
distance about 1 mm, called the external ring. (ii) The intervening
region between the internal and external rings are almost dark
except for some localized bright spots. (iii) Periodic bright spots
appear in the external ring. The bright spots follow the external
ring either when the excitation spot is moved over the sample, or
when the ring radius is varied with the excited power. (iv) PL is
washed out eventually with increasing temperatures. (v) In an
impurity potential well, the PL pattern becomes much more compact
than a Gaussian with a central intensity dip, exhibiting an annular
shape with a darker central region. (vi) With the increase of the
laser power, exciton cloud first contracts then expands. (vii)
Exciting sample by higher-energy lasers, the dip can turn into a tip
at the center of the annular cloud.

The above experimental facts raised several interesting questions
which need to be clarified. (i) What is the reason behind the
two-ring structure? (ii) What is the physical origin of the periodic
periodic bright spots in the external ring? And, (iii) is the a
coherent phenomenon, or a result due to unbalanced transportation?
In Ref.~\onlinecite{butov:117404}, a charge separated
transportation mechanism was proposed. It gave a satisfactory
explanation to the formation of the exciton ring and the dark region
between the internal and external rings. However, the remaining two
questions are still not clarified satisfactorily.

It is commonly believed that the periodic bright spots in the
external ring are formed due to certain kind of instability. It
generates the density modulation by a positive feedback. Regarding
the origin of the instability, Levite \emph{et al.} considered that
exciton states are highly degenerate and the instability comes from
the stimulated scattering.\cite{Levitov2004a} When a local
fluctuation occurs in the exciton density, it will lead to an
increase in the stimulated electron-hole binding rate.  The
depletion of local carrier concentration then causes neighboring
carriers to stream towards the point of fluctuation. On the other
hand, Sugakov suggests that the instability is due to an attractive
interaction between the high-density excitons.\cite{Sugakov2004a}
The creation of different structures, for example, the islands or
the rings of the condensed phase, occurs due to the nonequilibrium
state of the system connected with the finite value of the exciton
lifetime and the presence of pumping. Therefore, the appearance of
the patterns is likely a consequence of self-organization processes
in a nonequilibrium system. \cite{sugakov:115303}

To understand the subtle properties of indirect exciton, further
experiments have been done and reported. For example, a novel method
was proposed to demonstrate that cold exciton gases can be trapped
by the laser induced trapping, which is similar to the trapping of
BEC ultracold atoms. \cite{hammack:227402} More recently, an
improved trapping technique was used to lower the effective
temperature of indirect excitons. \cite{high-2008} With the even low
temperature, four narrow PL lines have been observed, which
corresponds to the emission of individual states of indirect
excitons in a disorder potential. The homogeneous line broadening
increases with density and dominates the linewidth at high
densities.

In a recent paper (Ref.~\onlinecite{yang:033311}), for the first
time, Yang \textit{et al.} measured the exciton PL energy along
the circumference of the ring. The most interesting result is that
the average exciton energy depends on temperature {\em
nonmonotonically}. With reducing temperatures, average exciton
energy of the indirect exciton is lowered until the transition
temperature, $T_{tr} \simeq 4$ K, is met. Below $T_{tr}$, the
macroscopically ordered exciton state is formed but the average
exciton energy in the ring actually increases with temperature
being further lowered. In particular, the largest energy of single
exciton is found to locate in the brightest regions. Due to these
observations, Yang \textit{et al.}\cite{yang:033311} argued that
the interaction between exciton is repulsive. A numerical
calculation seems to support this scenario.
\cite{schindler:045313}

Another important experiment is the measurement of the coherence
length. A Mach-Zehnder interferometer with spatial and spectral
resolution was used to probe the spontaneous coherence in cold
exciton gases, which are implemented experimentally in the ring of
indirect exciton in CQW. \cite{yang:187402} A strong enhancement of
the exciton coherence length is observed at temperatures below a few
Kelvin. The increase of the coherence length is correlated with the
macroscopic spatial ordering of excitons. The coherence length at
the lowest temperature corresponds to a very narrow spread of the
exciton momentum distribution, much smaller than that of a classical
exciton gas. It also shows that the apparent coherence length is
well approximated by the quadratic sum of the actual exciton
coherence length and the diffraction correction given by the
conventional Abbe limit divided by $\pi$. \cite{fogler:035411}

Whether the interaction between excitons in the macroscopically
ordered state is attractive or repulsive and to what extent the
interaction affects the formation of the macroscopically ordered
state remain an open question. The current paper attempts to
clarify the above issues. In fact, experimental data had revealed
several important clues. After analyzing the experimental facts,
it is found that attractive interaction is dominant in indirect
excitons and their nonequibirium distribution plays an important
role in the macroscopically ordered state. Moreover, the
uncertainty principle is shown to lead to a good account for the
experimental phenomena and numerical simulations have confirmed
that.

This paper is organized as follows. In Sec.~\ref{Temperature
dependence of exciton average energy in the exciton external ring
-- a qualitative analysis}, based on the uncertainty principle, we
present a qualitative analysis of the temperature dependence of
the average exciton energy in the external exciton ring. No
interaction is considered in this section. In Sec.~\ref{Particle
number density dependence of exciton PL FWHM broadening and energy
shift}, the uncertainty principle is used again to explain the
particle number density dependence of the PL FWHM (full width at
half maximum) broadening and energy shift. The PL spectra, which
are broadened first and then become sharper, provide a strong
evidence of the competition between a two-body attraction and a
three-body repulsion. In Sec.~\ref{Energy spatial distribution in
exciton external ring}, we use a phenomenological nonlinear
Schr\"{o}dinger equation, together with a temperature and energy
dependent exciton distribution, to discuss the exciton energy
spatial distribution in the external ring. The largest energy of
single exciton is found to be in the brightest regions. In
Sec.~\ref{Temperature dependence of exciton average energy in the
exciton external ring}, temperature dependence of the average
exciton energy in the external ring is studied quantitatively
using the approach proposed in Sec.~\ref{Energy spatial
distribution in exciton external ring}. The calculation shows that
average exciton energy first decreases and then increases with
increasing temperatures. In Sec.~\ref{Exciton number distribution
in laser induced trap}, we give a demonstration on the exciton
distribution in the laser induced trap case. Sec.~\ref{Summary} is
devoted to a brief summary.

\section{Temperature dependence of energy: qualitative analysis}

\label{Temperature dependence of exciton average energy in the
exciton external ring -- a qualitative analysis}

To understand the experimental phenomena, we first analyze the
electron and hole creation, transportation, and the exciton
formation qualitatively. As pointed out in
Refs.~[\onlinecite{butov:117404, rapaport:117405}], when electrons
and holes are excited by lasers, they are hot electrons and hot
holes initially. Since the drift speed of hot electrons is larger
than that of hot holes (electron has a smaller effective mass),
electrons and holes are indeed charge-separated at this stage (no
true exciton is formed). Because hot electrons and holes have a
small recombination rate and optical inactive, they can travel a very
long distance from the
laser spot. After a long-distance travel, hot electrons and holes
collide with the lattices and are eventually cooled down. Due to the
neutrality of the CQW, negative charges will slow down and
accumulate far away from the laser spot. As a consequence, spatially
separated electrons and holes form excitons and become optical active at the boundary of
the opposite charges where they recombine and show a sharp luminescence ring.
This kind of charge separated transportation mechanism has been used
to interpret the exciton ring formation.\cite{butov:117404,
rapaport:117405}

It should be emphasized that, however, after the long-distance
transportation and cooling, cooled electrons and holes will meet
in the region of the external ring in the experiment of Butov
\textit{et al.} \cite{Butov2002b}. However, in the experiment of
Lai \textit{et al.} \cite{LaiCW2004}, they meet in the impurity
potential well. It is believed that only cooled electrons and
holes can form excitons. At this stage, charges are not separated,
but coupled or bound together. The long lifetime of the excitons
means the lower electron-hole recombination rate. Therefore the
elemental particles in the external ring \cite{Butov2002b} and in
the impurity potential well \cite{LaiCW2004} are \emph{excitons}.
We thus argue that it is the interaction between excitons, rather
than the unbalanced transportation, which leads to the
nonhomogeneous density distribution associated with the complex PL
patterns.

When the experimental temperature is low ($T\alt 1.4$ K), average
translational kinetic energy and in-plane momentum of excitons are
also low. In such case, excitons are optically active and the
space-dependent PL intensity is proportional to the exciton number
distribution. Since exciton number distribution is equal to the
product of the probability distribution and the total excition
number $N$, if $N$ remains unchanged, PL intensity is directly
proportional to the probability distribution. In the following
discussion, we simply take exciton density distribution as the PL
distribution. Besides, since particles can only move within the
CQW, their motions are essentially two-dimensional (2D).

The periodic array of beads in the external rings is a
low-temperature phenomenon. This means that average exciton
translational kinetic energy is low enough and its wave-particle
duality is important. Therefore exciton energy is governed by the
uncertainty principle.
The uncertainty principle predicts that $\Delta p \sim \hbar/\Delta r$,
where $\Delta p$ is the momentum spread, $\Delta r$ is spatial
uncertainty, and $\hbar$ is the Planck's constant. Since exciton's
momentum can be approximated by $p\sim \Delta p$, the corresponding
energy $E\sim {\hbar^2}/[2m(\Delta r)^2]$. With these in mind, one is able to
consider the temperature dependence of the exciton energy
distribution, such as those reported in Ref.
\onlinecite{yang:033311}. When temperature is lower than
$T_{tr}\simeq 4$K, macroscopic ordered state forms. Exciton are
confined into each bead of the external ring. Their spatial
uncertainty $\Delta r$ can be approximated by the diameter $R$ of
the bead ($\Delta r \simeq R$). Experiments have observed that the
lower the temperature is, the higher contrast the pattern is and the
smaller the $R$ is. \cite{yang:033311} This implies that when
temperature is lower, momentum uncertainty will be larger and hence
the exciton energy will be higher. This gives a qualitative
interpretation why the average exciton energy increases with
decreasing temperatures.

When temperature is increased to be higher than $T_{tr}$,
macroscopically ordered state breaks down and the confined regions
enlarge essentially. The uncertainty related energy will decrease in
principle. However, when the temperature is increased further, the
system undergoes a transition into the classical limit to which
wave character of exciton is no longer important. In this case, the
average exciton translational kinetic energy is directly
proportional to the temperature.
Therefore, non-monotonic temperature dependence of the average
exciton energy is closely related to the pattern formation, and
has little to do with the interaction between excitons. In Fig.~3
of Ref.~\onlinecite{yang:033311}, linear temperature dependence
of the average exciton energy is reported at $T>T_{tr}$. Their
results confirmed that excitons behave classically in this limit
to which $E \sim k_{B}T$ at 2D. Nevertheless, the slope of the
linear temperature dependence was measured in Ref.
\onlinecite{yang:033311} to be about $1.4\times 10^{-4}$ eV
k$^{-1}$, slightly larger than the Boltzman constant,
$k_{B}=8.62\times 10^{-5}$ eV k$^{-1}$. The difference may be due
to some effect of the interactions and extra degrees of freedom in
addition to translational motions.

\section{Exciton PL spectra: qualitative analysis}
\label{Particle number density dependence of exciton PL FWHM
broadening and energy shift}

For a long time, indirect exciton PL FWHM energy and the shift
with increasing density in the presence of a random potential
remain to be fully understood.\cite{high-2008, butov:2004} In a
previous paper \cite{Liu2006}, we have proposed a mechanism to
study the macroscopically ordered exciton states. The most
important point lies in the interaction which involves the
competition between a two-body attraction and a three-body
repulsion. Here we elaborate this point, along with the
uncertainty principle, to illustrate the PL spectra.

It is instructive to emphasize several important features in
connection with our model. The idea first came from the density
dependence of inhomogeneous exciton distribution. It is clear that
the interaction between the indirect exciton is neither purely
attractive, nor purely repulsive. If the interaction between the
indirect exciton is purely repulsive, it will drive the exciton
towards homogeneous distribution and the exciton cloud will expand
with the increase of the exciton number. On the contrary, if the
interaction is purely attractive, the system is expected to collapse
when the exciton density is greater than a critical value to which
there is not enough kinetic energy to stabilize the exciton cloud.
In addition, the case of a purely repulsive or a purely attractive
interaction is incapable to understand the experimental fact that
the exciton cloud contracts first and expands later when the laser
power is increased.\cite{LaiCW2004}


The existence of the attractive interaction does not mean that the
exciton state is unstable against the formation of metallic
electron-hole droplet. The reason is that the repulsive
interaction may dominate over the attraction in that regime and
keep the system stable. Many effects may contribute to the exciton
interaction. The most important one is the exciton dipole-dipole
interaction. Exciton behaves like a dipole, so a strong repulsion
will govern exciton interaction when two dipoles are aligned
parallel. However, if the direction of two dipoles changes from
aligned parallel to inclined, the attraction between the election
of one exciton and the hole of the other exciton will dominate.
This case can happen easily when exciton density is low. In case
of high densities, due to the strong Coulomb interaction, the
dipoles will tend to align parallel and consequently a repulsive
interaction dominates. The other important interaction originates
from the exchange effect. When two indirect exciton approach to
each other, the exchange interaction between two electrons, as
well as the one between two holes become important. This may be
another source of the attractive interaction between excitons.

In fact, the complex exciton interaction which results from the
competition of multi-effects, may be well described by the van der
Waals form. It has pointed out that the effective interaction
between exciton will be attractive when the separation between two
exciton is about 3 to 6 exciton radii.\cite{Sugakov2004a} In the
current experiment, the exciton density is about $10^{10}/$
$\mathrm{cm}^{2}$. For this density, the average distance between
the indirect exciton is about 100 $\mathrm{nm}$. As the exciton
Bohr radius $a_{B}$ is about 10 $\sim$ 50 $\mathrm{nm}$
\cite{LaiCW2004}, the average distance between excitons is about
$2\sim10$ exciton radii. Thus it is reasonable to assume that the
two-body interaction is in the attractive regime. In fact, the
attractive interaction between the exciton has been considered as
a possible candidate to describe the pattern formation observed by
experiments. \cite{Sugakov2004a, Levitov2005}

In a previous paper\cite{Liu2006}, taking into account the
two-body attraction and the three-body repulsion interactions, we
have proposed that in the dilute limit the behavior of the exciton
can be described by a \emph{nonlinear} Schr\"{o}dinger equation
\cite{Liu2006}
\begin{equation}
-\frac{\hbar^2}{2m^{\ast}}\nabla^{2}\psi_{j}+(V_{ex}-g_{1}n+g_{2}n^{2})\psi
_{j}=E_{j}\psi_{j},  \label{the schrodinger equation}
\end{equation}
where $\psi_{j}$ and $E_{j}$ are the \textit{j-th} eigenstate and
eigenvalue, respectively. $V_{ex}$ is an external potential, and
$g_{1}$ and $g_{2}$ are (positive) coupling constants associated
with two-body and three-body interactions. $n=n({\bf r})$ is the
local density of exciton and at the mean field level,
\begin{equation}
n(\mathbf{r}) = \sum_{j=1}^{\mathcal{N}} \eta_{j}(E_{j})
|\psi_{j}(\mathbf{r} )|^{2},\label{eq:density}
\end{equation}
where $\mathcal{N}$ denotes the total number of bound
states and $\eta_{j}$ is the corresponding probability function
associated with the energy level $E_{j}$. $\eta_{j}$
satisfies the normalized condition $\sum_{j=1}^{\mathcal{N}%
}\eta_{j}(E_{j})=1 $. Readers can refer to
Ref.~\onlinecite{Liu2006} for a detailed description of the
model.

In the present model, when macroscopically ordered state forms in
the ring or in the impurity (disorder) potential well, exciton will
distribute over the discrete energy levels. In fact, four sharp
peaks in PL spectra corresponding to the emissions of indirect
exciton states have been observed recently in a so-called "elevated
trap".\cite{high-2008} We attribute the individual localized states
as the confinement due to both the elevated trap potential and the
self-trapped potential (arising from the attractive interaction).

The present model can give a natural and self-consistent
explanation of the particle density dependence of FWHM broadening
and PL energy shift. For low density exciton, the interaction
between exciton is dominated by the attraction. With the increase
of the particle density (by increasing the laser power), the low
temperature exciton cloud will contract irrespective of localized
or delocalized states. This means that the attractive interaction
hampers the exciton motion. At the mean field level, this
corresponds to the increase of the self-trapped potential. The
larger the particle density is, the larger the attractive
interaction is, and the exciton cloud will contract further. At
low temperature, $T < T_{tr}$, uncertainty principle governs the
exciton motion. Consequently, it results in higher mean exciton
energy and larger energy dispersion. Higher mean exciton energy
indicates blue-shift of PL peak and larger energy dispersion means
that PL peak is more broadened (FWHM becomes large).

When the particle density is further increased, three-body repulsion
becomes more important and self-trapped interaction becomes weaker.
The PL peaks will red shift and become sharp again. In fact, with
the increase of the laser power in the low particle density regime,
the phenomenon that PL spectra broaden first and then become sharper
has been observed (see Fig.~3(c) of Ref.~\onlinecite{high-2008}).
This, in addition to the support given by the experimental data in
Ref. \onlinecite{LaiCW2004}, provides a further evidence that the
interaction between exciton is likely the combination of a two-body
attraction and a three-body repulsion. With increasing the laser
power, blue shift of the PL peaks has been reported in Ref.
\onlinecite{Butov1990}. However, the prediction of blue shift
first and then red shift of the PL peak with the increase of the
laser power has not been observed yet! Further experiments are in
demand.


\section{Spatial energy distribution}

\label{Energy spatial distribution in exciton external ring}

In order to study the spatial dependence of the energy distribution
in the external exciton ring, it is necessary to solve the nonlinear
Schr\"{o}dinger equation (\ref{the schrodinger equation})
numerically. In addition to the assumption of interactions, another
important aspect lying in our model is the disequilibrium energy
disequilibrium energy distributions of indirect excitons. In fact,
the indirect excitons in CQW involves both the (complex) energy
relaxation and recombination processes. These have been studied by
several experimental and theoretical groups. \cite{Benisty1991} It
shows that for a relaxation process, when the exciton density is low
($n\ll a_{B}^{2}$, $a_{B}$ is Bohr radius), the effects due to the
exciton-exciton interaction and the exciton-carrier scattering can
be neglected. In this case, the relaxation time is mainly determined
by the scattering of excitons with acoustic phonons.
\cite{Piermarocchi1996} In particular, at low bath temperatures
$(T_{b}<1$ $\mathrm{K})$, this kind of relaxation rate decreases
dramatically due to the so-called "phonon bottleneck" effects.
\cite{Benisty1991}

For the recombination process, on the other hand, because exciton in
the lowest self-trapped level are quantum degenerate, they are
dominated by the stimulated scattering when the occupation number is
more than a critical value. Strong enhancement of the exciton
scattering rate has been observed in the resonantly excited
time-resolved PL experiment \cite{butov:5608}. Therefore, even
though the phonon scattering rate is still larger than the radiative
recombination rate, thermal equilibrium of the system may not be
reached. Essentially the distribution $\eta_{j}(E_{j})$ may deviate
significantly from the usual Boltzmann or Bose like.
\cite{Ivanov1999}

A recent paper (Ref.~\onlinecite{high-2008}) reported that the
observed narrow PL lines of indirect excitons in a disorder
potential correspond to the emission of individual states. It has
been shown that the intensity of PL spectra is roughly in
proportion to the exciton number. When excitons distribute over
some given energy levels, the PL spectra will exhibit strong peaks
corresponding to these levels. Therefore the relative height of
the PL peaks can be interpreted as the exciton distribution
probability associated with the energy level, $\eta_{j}(E_{j})$.

At the high temperature of 10 K, the PL peak intensity of the high
excited state is higher than that of the ground state (see Fig.~2(c)
of Ref.~\onlinecite{high-2008}). This indicates that exciton have
a larger probability distribution in the high excited state. With
reducing the temperature to 5.1 K, the PL peak intensity becomes
lower for the high excited state while it becomes higher for the
other three peaks associated with the two ground states. It seems
that excitons at higher states will relax to lower states when the
temperature is reduced. At the temperature lower than 5.1 K, the
intensities are found to be almost identical for the four peaks. It
indicates that excitons distribute over the four discrete levels
with almost the same probability. Furthermore, at even lower
temperature of 2.7 K, the low-energy peaks become very dim. This
implies that phonon bottleneck effect is in effect for ground state
and low excited states and in this regime, excitons in high excited
states are having difficulty relaxing to the ground states.

From the above analysis, one can generalize $\eta (E_{j})\rightarrow
\eta (E_{j},T)$ to have a better description for the temperature
dependence of particle number distribution of bound states.  As
mentioned above, experiment seems suggesting that the weight of
$\eta (E_{j},T)$ will transfer from high-energy states to low-energy
states when $T$ is decreased. It also suggests that, for a given
$T$, the weight of $\eta (E_{j},T)$ is larger in higher-energy
states than that in lower-energy states. With the above in mind, two
other factors are also important in determining the actual form of
$\eta (E_{j},T)$. Firstly, exciton are Bose quasiparticle and at not
too low temperatures, the distribution can be approximated by the
(classical) Boltzman function, $\exp (-E_{j}/T)$, where $T$ is
considered as an effective temperature related to lattice
temperature. The other important factor is the energy dependence of
exciton luminance efficiency. Low-energy excitons is more optical
active than that of high energy exciton. So the low energy excitons
have a high luminance efficiency which is contrary to the
high-energy excitons. Thus low-energy state will have relatively
smaller exciton distribution taking into account the luminance
efficiency. One can use $\exp (E_{j}/E_{0})$ to describe
qualitatively this effect, where $E_{0}$ is an effective energy
scale related to the exciton life time. As a consequence,
temperature and energy dependent exciton distribution is an outcome
of the competition between the above two effects. We then take the
following {\em phenomenological} form for the weight
\begin{eqnarray}
\eta (E_{j},T) &\equiv&C\exp \left[ -\left( E_{j}-\mu \right)
/T\right]
\exp \left[ \left( E_{j}-\mu \right) /E_{0}\right]    \nonumber\\
&=&C\exp \left[ \alpha (E_{j}-\mu)\right],   \label{eq:eta}
\end{eqnarray}
where $C$ is the normalization factor, $\mu$ is the chemical
potential, and $\alpha \equiv 1/E_{0}-1/T$. Typically $E_{0}<T$ and
hence $\alpha >0$. When temperature and laser power remain
unchanged, $\mu$ and $\alpha$ will remain constantly. While
increasing the temperature, $\alpha$ will increase correspondingly.
Thus $\alpha$ can be considered as an "effective temperature".

\begin{figure}[tbp]
\begin{center}
\includegraphics[width=8cm]{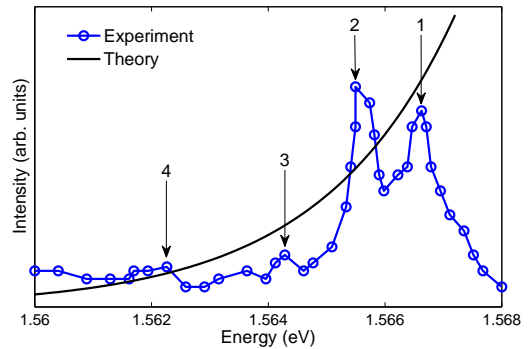}
\end{center}
\caption{Energy dependent exciton distribution. Solid line is
obtained from the phenomenological formula (\ref{eq:eta}) with
chemical potential $\mu=1.56$ eV and $\alpha=0.025$ eV$^{-1}$.
Circle line corresponds to the exciton PL spectra (in arbitrary
unit) taken from Fig.~2(a) of [\onlinecite{high-2008}]. For
comparison, a constant is multiplied to the solid line.}
\label{fig1}
\end{figure}

Equation (\ref{eq:eta}) basically gives an appropriate temperature
and energy dependent probability density distribution for indirect
exciton. In Fig.~\ref{fig1}, we compare the theoretical curve
[based on Eq.~(\ref{eq:eta})] with the experimental PL spectra
reported in Ref.~\onlinecite{high-2008}. The experimental data
(circles) in Fig.~\ref{fig1} are in arbitrary unit. For
comparison, we have multiplied a constant to the probability
density. It should be mentioned that we have tried various
distribution functions and found that Eq.~(\ref{eq:eta}) gives the
best fitting. In our earlier study (Ref. \onlinecite{Liu2006}), we
have assumed that exciton are distributed at different energy
levels with almost the same probability. Strictly speaking, it
corresponds only to the low temperature limit in Eq.
(\ref{eq:eta}).

Numerically it is convenient to do the following scaling:
$\psi_j(\mathbf{r})/\sqrt{N}\rightarrow \psi_j(\mathbf{r})$,
$Ng_{1}\rightarrow g_{1}$, and $N^{2}g_{2}\rightarrow g_{2}$, such
that Eq.~(\ref{the schrodinger equation}) remains the same form. In
this case, $n(\mathbf{r})$ becomes the probability
density which satisfies the normalization condition $\int _{S}n(\mathbf{r}%
)dS=1$. After further rescaling $\psi_j(\mathbf{r})\sigma_{\mathrm{PL}%
}\rightarrow \psi_j(\mathbf{r})$ and $\mathbf{r}/\sigma_{\mathrm{PL}%
}\rightarrow \mathbf{r}$, Eq.~(\ref{the schrodinger equation}) is reduced to
\begin{equation}
-\frac{1}{2}\nabla^{2}\psi_{j}+(v_{ex}-a_{1}n+a_{2}n^{2})\psi_{j}
=\varepsilon_{j}\psi_{j},  \label{the reduced schrodinger equation}
\end{equation}
where $v_{ex}\equiv V_{ex}/\epsilon$, $a_{1}\equiv g_{1}/\left( \sigma_{%
\mathrm{PL}}^{2}\epsilon\right)$, $a_{2}\equiv g_{2}/\left( \sigma_{\mathrm{%
PL}}^{4}\epsilon\right)$, and $\varepsilon_{j}\equiv E_{j}/\epsilon$. Here $%
\epsilon\equiv{\hbar^{2}}/{m^{\ast}} \sigma_{\mathrm{PL}}^{2}$ with $\sigma_{%
\mathrm{PL}}$ being the root-mean-square radius of the exciton
cloud observed by photoluminescence. With the above scaling, it is
found
that $a_{1}^{2}/a_{2}=g_{1}^{2}/g_{2}\epsilon=g_{1}^{2}{m^{\ast}} \sigma_{%
\mathrm{PL}}^{2}/g_{2}\hbar^{2}$, which is a constant for an explicit sample.

Taking into account the experimental facts,\cite{Butov2002b,
yang:033311} two important points should be clarified. (i) The
exciton patterns are fully determined by its self-trapped
interactions. External potential $V_{ex}$ is not the main cause for
complex exciton patterns. Thus we set $v_{ex}=0$ for simplicity.
(ii) When an electron and a hole form an exciton at low temperature,
it is believed that their kinetic energy is low and all exciton are
and all exciton are self-trapped. Particles with energy greater
potential energy can not be bounded in the self-trapped well and
should be ruled out in the calculations.

\begin{figure}[tbp]
\begin{center}
\includegraphics[width=8cm]{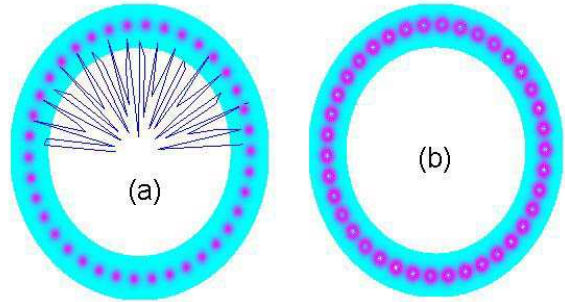}
\end{center}
\caption{(a) Formation of the macroscopically ordered state of the
exciton ring at very low temperatures. The corresponding
angle-dependent energy distribution $E_{k}(\theta)$ (solid line)
is also shown. (b) Spatial distribution of the kinetic energy,
$E_{k}(x,y)$. Parameters used are $a_{1}=250$,
$a_{2}=0.001a_{1}^{2}$, effective temperature $\alpha=0.002$, and
chemical potential $\mu=0$. } \label{fig2}
\end{figure}

The mean kinetic energy of excitons can be given by
\begin{eqnarray}
E_k =\int\int dx dy ~E_{k}(x,y), \label{1}
\end{eqnarray} where
$E_{k}(x,y)$ is the mean local (space-dependent) kinetic energy
density. Considering the probability function of each level,
$\eta_j\equiv \eta(E_j,T)$, $E_k$ can also be given by
\begin{eqnarray}
E_k =\sum_{j=1}^{\mathcal{N}} \eta_j E_{kj}, \label{2}
\end{eqnarray}
where $E_{kj}$ is the kinetic energy associated with level $j$. In
fact,
\begin{eqnarray}
E_{kj} &=&-\int\int dx dy ~\psi_j^*(x,y)\nabla^2\psi_j(x,y) \nonumber\\
&=& \int\int dx dy \left(\left|{\partial \psi_j(x,y)\over\partial
x}\right|^2+\left|{\partial \psi_j(x,y)\over\partial
y}\right|^2\right), \label{3}
\end{eqnarray} with $\psi_j(x,y)$ the wave function of level $j$.
In obtaining the 2nd line of Eq.~(\ref{3}), an integration by
parts and a boundary condition that wave function vanishes at the
infinity (bound states) are applied. By comparing
Eqs.~(\ref{1})--(\ref{3}), the mean spatial energy density is
obtained to be
\begin{eqnarray}
E_k(x,y)= \sum_{j=1}^{\mathcal{N}} \eta_j \left(\left|{\partial
\psi_j(x,y)\over\partial x}\right|^2+\left|{\partial
\psi_j(x,y)\over\partial y}\right|^2\right). \label{4}
\end{eqnarray}
Besides, one can also obtain the angle dependent kinetic energy
density via
\begin{equation}
E_{k}\left( \theta \right) =\int\limits_{\left( x,y\right) \in
\left(\theta,\theta +\Delta \theta \right) }dxdy~ E_{k}(x,y)
\end{equation}%
with $\tan \theta \equiv y/x$. In Fig.~\ref{fig2}(a), we show both
the particle density distribution, $n(\mathbf{r})$, in the exciton
ring and the angle-dependent kinetic energy distribution,
$E_{k}\left( \theta \right)$. Obviously the maxima of $E_{k}\left(
\theta \right) $ are located at the center of the beads.
Fig.~\ref{fig2}(b) shows the spatial dependence of the kinetic
energy distribution, $E_{k}(x,y)$. One interesting feature is that
$E_{k}(x,y)$ is zero at the center of the beads. This can be
elucidated by future experiments of finer resolution.

\section{Temperature dependence of energy: quantitative results}

\label{Temperature dependence of exciton average energy in the exciton
external ring}

\begin{figure}[ptb]
\begin{center}
\includegraphics[width=8cm]{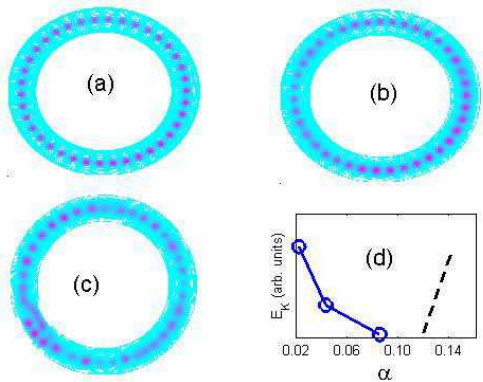}
\end{center}
\caption{Formation of the macroscopically ordered state of the
exciton ring at different effective temperatures: (a)
$\alpha=0.002$, (b) $\alpha=0.004$, and (c) $\alpha=0.008$. (d)
Temperature ($\alpha$) dependence of the total energy $E_{k}$. All
other parameters are taken to be the same as those in
Fig.~\ref{fig2}.} \label{fig3}
\end{figure}

In section \ref{Temperature dependence of exciton average energy
in the exciton external ring -- a qualitative analysis}, with the
uncertainty principle, a qualitative analysis was made on the
temperature dependence of exciton average energy in the external
ring. Here we solve the nonlinear Schr\"{o}dinger equation
(\ref{the schrodinger equation}), together with the disequilibrium
distribution $\eta_j(E_j,T)$, to study the temperature dependence
quantitatively. As mentioned before, $\alpha$ can be treated as an
effective temperature in the present case. Temperature
($\alpha$)-dependent exciton ring patterns are shown in
Fig.~\ref{fig3}(a)-(c), while Fig.~\ref{fig3}(d) shows the
nonmonotonic temperature dependence of the total energy,
$E_{k}=\int E_{k}(\theta) d\theta$.

It should be noted that our theory applies only to low
temperatures when excitons are in or near the condensed state.
When temperature is above the critical point, such as
$T_{tr}\simeq 4$ K presented in Ref.~\onlinecite{high-2008},
excitons will transfer from condensed to non-condensed phase.
Above $T_{tr}$ in the non-condensed phase, the exciton liquid may
be ``boiling" and its spatial distribution could change with the
time. At such high temepratures, our theory will break down to
which numerical calculation turns no convergent solutions. In
fact, the exciton pattern shown in Fig.~\ref{fig3}(c) is just a
snapshot of time-dependent exciton number distribution. When
$\alpha\agt 0.08$ in our numerical simulation, the exciton
behavior can be well described by the classical statistical
physics and a linear formula $\varepsilon=k_B T$ is plotted in
Fig.~\ref{fig3}(d) for illustration purpose.

A brief summary is in order here. Initially, if exciton are
uniformly distributed on the external ring, attractive interaction
will drive the exciton to approach each others. When the local
density reaches a critical value, the kinetic energies of the
exciton will drive the high-density excitons to diffuse. At the same
time, the repulsive interaction, which increases with the density,
will hinder further increase of the exciton density. As a result,  
an array of clusters on the external ring forms. The size of
these clusters is thus determined by three factors, i.e., the
attractive interaction, the three-body repulsive interactions, and
the kinetic energies of the excitons associated with the
temperatures. This is the basic idea behind our model on the
macroscopically ordered states of excitons.

Finally some remarks on the temperature effect are given. When the
bath temperature is low, cooled exciton have relatively low momenta
and the self-trapped interaction is able to confine most of the
excitons. However, in the low momentum case, cooling efficiency is
low while luminous efficiency is high, excitons can not reach the
thermal equilibrium state. At the meantime, due to the competition
between the self-trapped and the kinetic energies, complex exciton
patterns occur (as discussed above). With increasing the
temperature, the exciton can not be fully cooled and correspondingly
self-trapped interaction confines only part of the excitons. The
attractive interaction also cannot compensate the exciton kinetic
energy and excitons will distribute homogenously in a 2D plane. In
this case, the pattern is washed out. If the temperature is higher
than the indirect exciton binding energy $\sim3.5$ $\mathrm{meV},$
\cite{Snoke2004b,szymanska:193305} most of excitons become ionized
and are in a plasma state. No pattern can be observed in this case.

\section{Number distribution with laser induced trap}

\label{Exciton number distribution in laser induced trap}

Similar to the studies of ultracold alkali atoms and molecules,
the laser-induced trapping was proposed and demonstrated for a
highly degenerate Bose gas of exciton. \cite{hammack:227402} An
important advantage of laser trapping is that it is possible to
control the trap in situ by varying the laser intensity in space
and time. A commonly seen phenomenon for highly degenerate
excitons is its annual particle distribution in the laser-induced
trap. We argue that this kind of distribution is similar to that
found in an impurity potential reported in Ref. \onlinecite{LaiCW2004}. 
In our previous
work, a detailed discussions was already given to this kind of
trapped exciton distribution.\cite{Liu2006} Since laser induced
trapping can confine higher density exciton, it is hoped that the
experiment can also be done at very high densities to explore the
many-body physics of excitons.

\begin{figure}[ptb]
\begin{center}
\includegraphics[width=9cm]{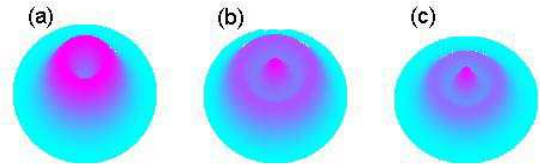}
\end{center}
\caption{The exciton distribution in laser-induced trap for (a) low
($a_1=15$), (b) intermediate ($a_1=23$), and (c) high ($a_1=35$)
densities. The other parameters are taken to be same as those in
Fig.~\ref{fig2}.} \label{fig4}
\end{figure}

Using the temperature and energy dependent distribution given in
Eq.~(\ref{eq:eta}) together with a two-dimensional pseudopotential
\begin{equation}
v_{ex}(\mathbf{r}) =\left\{
\begin{array}{cc}
\displaystyle -5 ~~ & \mathrm{for}~~r\leq \sigma_{\mathrm{PL}} \\
0 & \mathrm{otherwise,}%
\end{array}
\right.  \label{V}
\end{equation}
we solve the nonlinear Schr\"{o}dinger equation (\ref{the
schrodinger equation}) for exciton number distribution ranging
from low to high densities. The results are presented in
Fig.~\ref{fig4}. As mentioned before, $\sigma_{\mathrm{PL}}$ is
the root-mean-square radius of the exciton PL pattern. To compare
to experiments, we take $a_{1}$ = $15$, $23$, and $35$ with
$a_{2}=0.004a^{2}_{1}$ accordingly. As shown in Fig.~\ref{fig4},
the excitons distribute annularly at low densities and at high
densities, some fine structures develop at the center of the ring,
which is consistent with the experiments. Readers can refer to
Ref.~\onlinecite{Liu2006} for the reason behind the annular
distribution and some related physics.

\section{Summary}

\label{Summary}

In summary, the uncertainty principle is used to analyze the spatial
and temperature dependence of the average exciton energy
distribution in a macroscopically ordered state of the exciton. the
competition between a two-body attraction, a three-body repulsion,
and the kinetic energy plays a crucial in determining the behaviors
of exciton distributions. Numerical simulation of the corresponding
nonlinear Schr\"odinger equation seems to confirm the analysis.
Nevertheless, the reason of forming macroscopically ordered exciton
states may be more complex than what is expected. Full understanding
of the exciton interactions requires a reliable many-body
calculation beyond the mean-field approximation, which is obviously
not feasible at the moment.
In order to realize the exciton BEC phenomenon, it is in demand to
obtain the exciton with even lower combination rate and shorter
relaxation time.

\begin{acknowledgments}
This work was supported by the 
National Basic Research Program of China (Grant No.
2005CB32170X), and National Science Council of Taiwan (Grant No.
96-2112-M-003-008).
\end{acknowledgments}


\end{document}